\documentclass[11pt]{article}
\usepackage[utf8]{inputenc}
\usepackage[T1]{fontenc}
\usepackage{lmodern}
\usepackage[margin=1in]{geometry}
\usepackage{amsmath,amssymb,booktabs}
\usepackage{hyperref}
\usepackage{xcolor}
\hypersetup{colorlinks=true,linkcolor=blue,citecolor=blue,urlcolor=blue}

\title{False Security Confidence in Benign LLM Code Generation\thanks{Repository: \url{https://github.com/SysSoftwareSecLab/fsc-benchmark}}}
\author{%
  Xiaolei Ren\thanks{Correspondence: \texttt{xlren@must.edu.mo}} \\
  \small School of Computer Science and Engineering \\
  \small Macau University of Science and Technology \\
  \small Taipa, Macao SAR, China
}
\date{April 2026}

\begin{document}
\maketitle

\noindent\textbf{Keywords:} false security confidence, FSC rate, LLM code generation, security evaluation, functional correctness, benchmark design, deployment-context tasks, FSC-hard, conditional security metric

\begin{abstract}
Prior work has demonstrated that functionally correct yet vulnerable outputs arise systematically in threat-oriented settings, where adversarial or implicit channels are used to induce security failures in code agents and automated patching workflows~\cite{fcv2025}. This note introduces a complementary but distinct framing: \textbf{False Security Confidence (FSC)}, which studies the same surface phenomenon from a \emph{measurement-first perspective} in ordinary, non-attack-framed generation tasks. Our interest is not in whether attacks can produce such outputs, but in how frequently and in what forms they appear \emph{absent explicit attack pressure}, and whether conventional functional evaluation reliably detects them. We formalize \textbf{FSC rate} as the prevalence of security failure \emph{within the set of functionally correct outputs}, distinguish it from prior joint functional--security metrics such as SAFE and outcome-driven evaluation frameworks such as CWEval~\cite{safe2025,cweval2025}, define a three-ecosystem task view for studying how FSC manifests across general-purpose programming, deployment-context tasks, and security-explicit programming, and identify \textbf{FSC-hard} as a practically important refinement layer in which static analyzers miss vulnerabilities that remain dynamically triggerable. This technical report is intentionally scoped as a framework statement rather than a full empirical paper: its purpose is to establish terminology, measurement boundaries, and study design commitments for subsequent large-scale evaluation.
\end{abstract}

\section{Introduction}
Functional correctness and security correctness are often treated as if they were naturally aligned in LLM-based code generation. In practice, however, they frequently diverge. A generated program may satisfy unit tests, produce the expected output, or even look stylistically polished, while still embedding unsafe query construction, insecure credential handling, exploitable deserialization logic, or context-sensitive deployment misconfigurations. Recent evaluation work already shows that this mismatch is not rare: early evidence from formally published studies of AI-assisted code generation security already documented insecure suggestions in realistic development settings~\cite{pearce-copilot2022}, and more recent secure-code benchmarks and joint evaluation studies repeatedly report a non-trivial share of outputs that appear functionally successful while remaining security-problematic~\cite{cweval2025,safe2025,safegenbench2025}. This gap creates a particularly misleading form of evaluation comfort: once a sample is counted as ``correct,'' both users and researchers may implicitly overestimate its quality. We refer to this phenomenon as \emph{False Security Confidence (FSC)}: confidence induced by functional success even though the generated artifact remains security-failing. The goal of this note is not to claim discovery of the broader functionally-correct-yet-vulnerable phenomenon, which has already been observed in adjacent lines of work, but to define a clean \emph{measurement framework} for studying how frequently and in what forms this mismatch appears during ordinary, non-attack-framed LLM-assisted programming.

\paragraph{Contributions of this note.}
This report makes four scoped contributions. First, it introduces FSC as a measurement-oriented concept for studying security failure \emph{within} functionally correct outputs. Second, it formalizes FSC rate so that security failure is normalized by the set of functionally correct outputs rather than by all generations. Third, it proposes a three-ecosystem view of tasks that separates general-purpose programming, deployment-context tasks, and security-explicit programming. Fourth, it defines FSC-hard, a refinement layer for cases that evade static tooling yet remain dynamically exploitable.

\section{Positioning FSC Relative to Prior Work}
FSC should be read as a \emph{measurement-first reframing}, not as a competing claim that functionally correct yet vulnerable code was never previously identified. Prior work has already shown that code agents and code generation systems can produce outputs that preserve functionality while embedding vulnerabilities, spanning both foundational studies of AI coding assistants in practice~\cite{pearce-copilot2022} and more recent settings that motivate adversarial or threat-oriented analysis~\cite{fcv2025}. Our claim is narrower and different: even when the interaction is not framed as an attack, and even when the user's visible objective is simply to solve a programming task, the evaluation pipeline can still produce the same misleading end state---a functionally successful artifact that should not be trusted from a security standpoint. The scientific value of FSC therefore lies in changing the \emph{measurement lens}: from threat induction and exploitability demonstrations to the prevalence, distribution, and structural causes of security failure \emph{inside the subset of outputs that conventional evaluation would otherwise count as success}.

\paragraph{Relationship to SAFE and joint evaluation work.}
Recent work has already argued that secure code generation should not be evaluated on security and functionality in isolation. SAFE introduces a joint functional--security metric for comparing secure code generation methods, and CWEval provides an outcome-driven framework that tests both aspects on the same problem set~\cite{safe2025,cweval2025}. FSC is complementary rather than redundant. SAFE is a \emph{method-comparison metric}: it asks how well a secure-code-generation technique performs when security and functionality are jointly considered. FSC rate is a \emph{conditional failure metric}: it asks how often security failure occurs \emph{among the outputs that would already be counted as functionally correct}. The denominator is therefore the key distinction. FSC is designed to quantify over-trust risk inside apparent success, not to collapse multiple desiderata into a single overall score.

\paragraph{Relationship to internal-mechanism work.}
Recent work has shown that code LLMs can internally represent security concepts even while generating insecure code~\cite{scscode2026}, suggesting that FSC is not merely a capability gap but may also involve a generation-priority divergence. FSC as a measurement framework is complementary to this finding: rather than explaining why models generate insecure-yet-correct code, FSC quantifies how often this occurs in ways that escape conventional evaluation.

\paragraph{Boundary statement.}
Under this framework, FSC is \textbf{not} an umbrella term for repair-time phenomena, optimization-time phenomena, or analyzer disagreement. It is strictly reserved for initial-generation settings in which the output is functionally correct but security-failing. Apparent repair phenomena that arise when users attempt to remediate FSC instances through conversational refinement belong to a separate taxonomy of Pseudo-Repair~\cite{pseudorepair-note}, which is treated as a companion framework to the present note.

\section{Definition of False Security Confidence}
\paragraph{Notation.}
Throughout this note, $s$ denotes a single generated code sample produced by a fixed model under a fixed prompt, $\mathrm{FCR}(s) \in \{0,1\}$ is the functional correctness indicator under the task's oracle, and $\mathrm{SSR}(s) \in \{0,1\}$ is the security success indicator under the chosen validation stack. Both indicators are treated as deterministic given the sample and the validation procedure; stochasticity enters through the generation distribution, not through the indicators.

We define FSC at the sample level. A sample exhibits \textbf{False Security Confidence} if and only if
\[
\mathrm{FCR}(s)=1 \;\wedge\; \mathrm{SSR}(s)=0.
\]
Intuitively, FSC marks the subset of outputs that are likely to be over-trusted because they ``work'' in the narrow functional sense while remaining unsafe in the broader engineering sense.

\paragraph{Interpretation.}
The definition intentionally puts functionality first because the problem we want to quantify is not generic vulnerability prevalence, but rather the vulnerability prevalence \emph{among outputs that would otherwise be counted as successful generations}. This is what differentiates FSC from ordinary insecure-code rate and makes it especially relevant for benchmark design, deployment decisions, and scientific claims about model quality.

\section{FSC Rate as a Conditional Security-Failure Metric}
We define \textbf{FSC rate} over a set of generations $S$ as
\[
\mathrm{FSC\mbox{-}rate}(S)=
\frac{\left|\left\{s \in S : \mathrm{FCR}(s)=1 \wedge \mathrm{SSR}(s)=0\right\}\right|}
{\left|\left\{s \in S : \mathrm{FCR}(s)=1\right\}\right|}.
\]
This conditional denominator is essential. If one instead divides by all generations, then changes in general coding ability and changes in security behavior become confounded. FSC rate isolates the portion of \emph{apparent success} that should be discounted because the success is security-invalid. As a result, two models can have similar overall vulnerability rates but very different FSC profiles if one produces many more functionally correct outputs than the other. FSC rate is defined only when the denominator is non-zero; tasks or model-task pairs yielding no functionally correct outputs are excluded from FSC analysis by definition.

\paragraph{Reporting recommendation.}
We recommend that FSC rate never be reported alone. At minimum it should be paired with functional correctness rate, and when possible with raw insecure-code rate, analyzer coverage notes, and a small set of manually adjudicated exemplars. The key interpretive question is not just ``how often is the model insecure?'' but ``how often is the model insecure \emph{precisely when evaluators are most tempted to trust it}?''

\section{A Three-Ecosystem View of FSC}
We propose studying FSC across three task ecosystems because the causes of misleading security success are unlikely to be uniform. \textbf{General-Purpose Programming} includes common algorithmic, input/output, and data-processing problems with no explicit security framing; here FSC is expected to arise when latent vulnerabilities are masked by narrow task correctness. \textbf{Deployment-Context Tasks} introduce secrets handling, authorization logic, request routing, configuration semantics, environment assumptions, and service boundaries; here functionally correct code may still fail under realistic deployment or threat assumptions. \textbf{Security-Explicit Programming} directly requests validation, cryptographic usage, sanitization, or access control; here FSC captures the disturbing case where the model appears to understand the security goal yet still produces an implementation that remains vulnerable. Existing benchmarks have broadened secure-code evaluation coverage, but they do not organize the space around this measurement question or around the deployment-context failure modes FSC is intended to expose~\cite{safegenbench2025}. In particular, benchmark suites typically foreground prompt-level task categories or method-level comparison settings, while under-specifying the environment- and deployment-dependent conditions under which security relevance only becomes visible after the code is situated in a realistic execution context.

\paragraph{Why the ecosystem split matters.}
This partition is not merely organizational. It is a hypothesis about \emph{where} different forms of false confidence are most likely to arise. A task that appears harmless in a general benchmark may conceal security relevance once embedded in a deployed environment. Conversely, a security-explicit task may reveal whether the model can maintain security intent under pressure rather than merely produce plausible secure-looking syntax.

\section{FSC-hard as a Refinement Layer}
Not all FSC instances are equally visible. Some are caught by standard static analyzers; others are not. We define \textbf{FSC-hard} as the subset of FSC cases in which static tooling fails to flag the vulnerability while lightweight dynamic or semantic validation still confirms that the exploit path remains reachable. FSC-hard is a \textbf{refinement layer} within the FSC framework, not a separate top-level problem. Its purpose is to identify the portion of FSC instances that are most resistant to current tooling---precisely the cases that practitioners are least likely to catch without dedicated dynamic validation.

\paragraph{Why FSC-hard matters.}
If a functionally correct yet insecure output is at least statically flagged, users still receive a visible warning. FSC-hard is more dangerous because it aligns apparent success with apparent tool silence. This makes it the most practically deceptive region of the FSC space. It is also where current evaluation stacks are most likely to undercount risk, particularly when static analysis is treated as the dominant or sole security oracle~\cite{safe2025,scscode2026}.

\section{Validation Guidance for Future Empirical Work}
Because FSC is a measurement framework, its empirical value depends on validation discipline. Functional correctness should be established through task-specific oracles rather than stylistic judgment. Security success should be evaluated through a layered stack that can include multiple static analyzers, dynamic tests where feasible, and selective manual adjudication for disagreement-heavy or high-impact cases. The exact stack may vary by ecosystem, but the principle should remain fixed: FSC claims require enough evidence to support the statement that an output is both functionally successful and security-failing.

\paragraph{Minimal credible validation principle.}
This note does not prescribe one universal maximum-cost validation stack. Instead, it recommends the minimum stack sufficient for the research question. For general FSC measurement, paired functional and security evaluation may suffice. For FSC-hard claims, however, dynamic or semantic evidence is necessary by definition. Specifically, a dynamic exploit oracle or a concrete exploit-path demonstration constitutes sufficient evidence; warning disappearance or code-pattern matching alone does not.

\section{Scope and Non-Goals}
This report does not present prevalence estimates, benchmark results, or a fixed leaderboard. It does not claim that FSC replaces existing secure-code-generation metrics, nor that it subsumes adversarial threat analyses such as FCV. It also does not resolve the internal causal question of why models generate insecure-yet-correct outputs; that question belongs to mechanistic work such as SCS-Code~\cite{scscode2026}. Apparent repair phenomena that arise when users try to remediate FSC instances through conversational refinement belong to a separate taxonomy of Pseudo-Repair~\cite{pseudorepair-note}, which should be read as a companion framework rather than a subsection of FSC.

\section{Conclusion}
False Security Confidence names a measurement problem that has been visible in fragments across recent secure-code-generation research but has not, to our knowledge, been cleanly isolated as a conditional security-failure concept. By defining FSC, FSC rate, a three-ecosystem task view, and FSC-hard, this note fixes a vocabulary for studying how conventional notions of ``correctness'' can mask security failure in benign LLM code generation. Planned large-scale empirical evaluation will quantify FSC rate across models, languages, and task ecosystems; the companion framework addressing repair-stage failure modes is described in~\cite{pseudorepair-note}. Planned benchmark studies are documented at the repository listed above.

\section*{Acknowledgments}
This work was supported by the Science and Technology Development Fund (FDCT) of Macau SAR under Grant No.~0011/2025/ITP1.

\end{document}